\documentclass[conference]{IEEEtran}
\IEEEoverridecommandlockouts
\usepackage{cite}
\usepackage{amsmath,amssymb,amsfonts}
\usepackage{algorithmic}
\usepackage{graphicx}
\usepackage{wrapfig}
\usepackage{textcomp}
\usepackage{soul}
\usepackage{xcolor}
\usepackage[a4paper, total={184mm,239mm}]{geometry}
\def\BibTeX{{\rm B\kern-.05em{\sc i\kern-.025em b}\kern-.08em
    T\kern-.1667em\lower.7ex\hbox{E}\kern-.125emX}}

\title{COFFEE: A Carbon-Modeling and Optimization Framework for HZO-based FeFET eNVMs\\

}

\author{
    \IEEEauthorblockN{
    Hongbang Wu\IEEEauthorrefmark{1}, 
    Xuesi Chen\IEEEauthorrefmark{2}, 
    Shubham Jadhav\IEEEauthorrefmark{1}, 
    Amit Lal\IEEEauthorrefmark{1}, 
    Lillian Pentecost\IEEEauthorrefmark{3},
    and Udit Gupta\IEEEauthorrefmark{2}}
    
    \IEEEauthorblockA{\IEEEauthorrefmark{1}Cornell University, 
    \IEEEauthorrefmark{2}Cornell Tech, 
    \IEEEauthorrefmark{3} Amherst College}

    \IEEEauthorblockA{
    \{hw773, xc562, saj96, amit.lal, ugupta\}@cornell.edu, lpentecost@amherst.edu}

    \vspace{-10mm}
}

\begin{document}

\maketitle

\begin{abstract}
Information and communication technologies account for a growing portion of global environmental impacts.
While emerging technologies, such as emerging non-volatile memories (eNVM), offer a promising solution to energy efficient computing, their end-to-end footprint is not well understood.
Understanding the environmental impact of hardware systems over their life cycle is the first step to realizing sustainable computing.
This work conducts a detailed study of one example eNVM device: hafnium–zirconium-oxide (HZO)-based ferroelectric field-effect transistors (FeFETs).
We present \textbf{COFFEE}, the first carbon modeling framework for HZO-based FeFET eNVMs across life cycle, from hardware manufacturing (embodied carbon) to use (operational carbon).
COFFEE builds on data gathered from a real semiconductor fab and device fabrication recipes to estimate embodied carbon, and architecture level eNVM design space exploration tools to quantify use-phase performance and energy. 
Our evaluation shows that, at 2\,MB capacity, the embodied carbon per unit area overhead of HZO-FeFETs can be up to 11\% higher than the CMOS baseline, while the embodied carbon per MB remains consistently about 4.3$\times$ lower than SRAM across different memory capacity.
A further case study applies COFFEE to an edge ML accelerator, showing that replacing the SRAM-based weight buffer with HZO-based FeFET eNVMs reduces embodied carbon by 42.3\% and operational carbon by up to 70\%.


\end{abstract}


\section{Introduction}
Recent studies indicate that information and communication technology sector currently accounts for approximately 1.8\%-2.8\% of global greenhouse gas (GHG) emissions~\cite{1}, and projections suggest that this share could increase several-fold within the next decade~\cite{2}.
Recent studies have shown that as operational energy is increasingly supplied by renewable sources, embodied carbon from hardware manufacturing processes, including electricity consumption during fabrication and GHG emissions, has become the dominant contributor to the life cycle carbon impact~\cite{3,4}. 
Developing accurate models to quantify and optimize computing hardware carbon emissions  is crucial for guiding sustainable system design and deployment.
 
Embedded non-volatile memories (eNVMs), such as resistive RAM (RRAM) and ferroelectric field-effect transistors (FeFETs), are promising technologies to provide high-density, energy-efficient on-chip memory~\cite{8}. 
By comparison, conventional CMOS SRAM remains the mainstream on-chip memory but incurs substantial overheads.
Recent state-of-the-art (SOTA) accelerator shows that 6T SRAM accounts for 67\% of the chiplet area and contributes 56\% of the operational energy consumption~\cite{6,7}. 
In contrast, eNVMs combine compact cell structures with non-volatility, allowing data retention without high leakage power and delivering substantial density and energy-efficiency advantages~\cite{9,10}. 
These benefits have led to extensive investigation in academia~\cite{NVMExplorer} and adoption in consumer products~\cite{36}, providing competitive solutions despite potential drawbacks of comparatively higher write latency, variable reliability and scaling, and limited endurance~\cite{8,11, NVMExplorer}. 
However, the sustainability impact of eNVMs has not yet been considered.  
As the integration of eNVM with standard CMOS may require modified fabrication steps and additional materials, these changes potentially increase the energy intensity of production and thus may incur higher embodied carbon emissions per wafer~\cite{19}. 

\begin{figure}[t]
    \centering
    \includegraphics[width=\columnwidth]{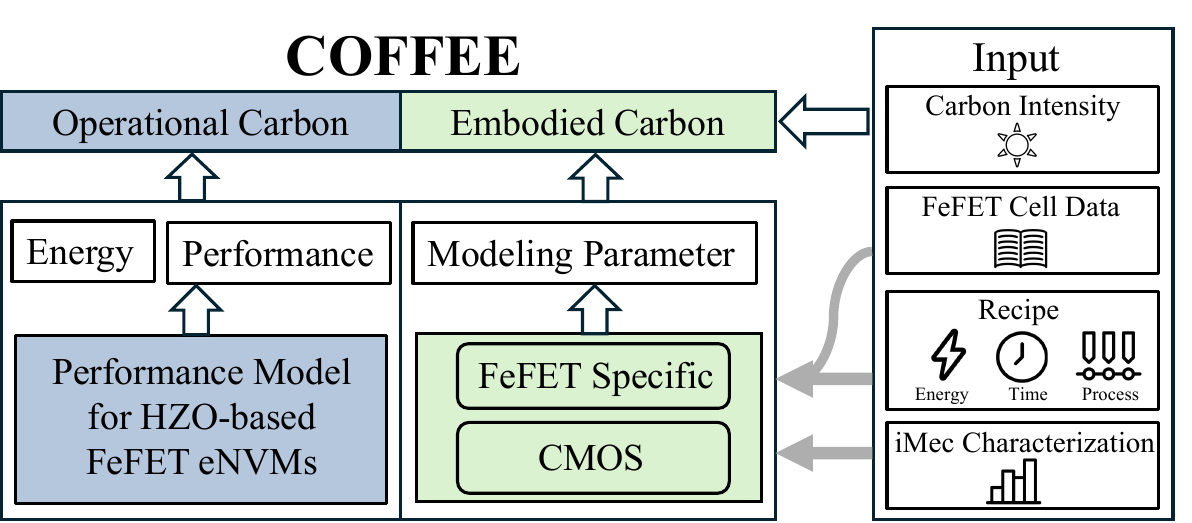} 
    \vspace{-1.2\baselineskip}
    \caption{COFFEE framework encompasses both embodied carbon and operational carbon. The embodied carbon from FeFET fabrication is partitioned into a CMOS baseline, evaluated using ACT~\cite{4}, and a FeFET-specific component leveraging real manufacturing recipes that provide tool energy and tool time. The operational carbon is obtained from performance and power evaluation using NVMExplorer~\cite{NVMExplorer}. Together, the two stages capture both manufacturing and execution, providing a life cycle analysis on the carbon footprint for FeFET.}
    \vspace{-1.3\baselineskip}
    \label{fig1}
\end{figure}

Although numerous tools have been developed to evaluate and optimize the energy efficiency of eNVM-based circuits and systems~\cite{12, NVMExplorer}, they focus primarily on technology characterization and performance metrics during use. 
Such analyses fail to capture the sustainability impact of fabrication and operation. 
Developing such a life cycle carbon analysis for eNVMs faces several challenges. 
First, details of the additional fabrication steps required for integrating specialized eNVMs are unknown or unreported in existing publications, particularly salient data to evaluate the manufacturing energy demand and GHG emissions. 
Second, the model should reflect cutting-edge device characteristics and process parameters, such as material layer thickness, to estimate carbon emissions at current technology nodes. 
Finally, the limited endurance of eNVM devices introduces reliability concerns and limits on device lifetime that must be incorporated into sustainability assessments. 

In order to demonstrate an end-to-end life cycle carbon analysis of a competitive eNVM technology, we leverage  detailed fabrication recipes for hafnium–zirconium-oxide (HZO) FeFETs, a highly CMOS-compatible and scalable eNVM device that has become particularly competitive and widely adopted in recent research~\cite{20}. To quantify the energy, gas, and material impact of HZO fabrication, we gather process descriptions for atomic layer deposition (ALD) steps together with tool power measurements directly from a semiconductor fab~\cite{Recipe}. In this context, recipe specifically refers to the real fab-based data sets. These data enable quantitative estimation of  embodied carbon associated with FeFET-specific fabrication steps.


We introduce a carbon model that includes embodied carbon and operational carbon, COFFEE, for HZO FeFETs (Figure~\ref{fig1}). 
The modeling flow separates the total embodied carbon into two parts: a CMOS baseline computed with ACT~\cite{4}, which relies on iMec characterization~\cite{32}, and a FeFET-specific component. 
For the latter, the model provides a step-by-step analysis of fabrication stages and corresponding carbon footprint impact from recipes and quantifies process-specific parameters, thereby offering clear insights for device and technology designers.
Furthermore, we collect recent publications on HZO-based FeFETs to extract parameters for lifetime and embodied carbon analysis~\cite{21,22,23,24,25,26,27}. 
Finally, we employ NVMExplorer~\cite{NVMExplorer} to evaluate operational parameters such as memory array latency and energy, enabling a systematic discussion of the trade-off between operational carbon and embodied carbon across diverse architectural configurations and application workloads.The main contributions of this work are:

\begin{enumerate}
    \item  We propose a framework (COFFEE) for evaluating the embodied carbon emissions of HZO-based FeFET  eNVMs. COFFEE models the fabrication process, with detailed step-level analysis and configurable parameters according to emerging device characteristics.

    \item We present a comparative analysis of embodied  carbon emissions between HZO-based FeFETs vs. conventional SRAM at a fixed storage capacity of 2\,MB. Under the same optimal-target configuration. The embodied carbon overhead of HZO-FeFETs per unit area can be as high as 11\% compared to CMOS baseline. We also compare representative HZO-FeFETs and SRAM across capacities from 2\,MB to 32\,MB, the embodied carbon per MB remains consistently around 4.3$\times$ better than SRAM, highlighting the significant sustainability advantage achievable through high memory density.

    \item Using COFFEE, we study the carbon impact of implementing edge AI accelerators using HZO-FeFETs. The case study shows 42.3\% embodied carbon reduction and up to 70\% operational carbon reduction per inference.

\end{enumerate}

COFFEE is available in the below GitHub repository: https://github.com/S4AI-CornellTech/COFFEE

\section{Background and Related Works}

In this section, Section~\ref{sec:carbon_modeling} overviews carbon modeling methodologies and related work, Section~\ref{sec:fefet_background} summarizes the advantages and challenges of FeFET technologies with an emphasis on HZO-based FeFET, and Section~\ref{sec:model_limitations} discusses the limitations of existing FeFET eNVM models.

\subsection{Carbon modeling} 
\label{sec:carbon_modeling}
The total carbon footprint (CF) of a computing system is defined as the sum of its operational (OCF) and embodied carbon footprints (ECF), with the embodied carbon proportion according to the fraction of application run‑time(T) over the chip expected lifetime (LT).
\vspace{-0.3\baselineskip}
\begin{equation}
\mathrm{CF} = \mathrm{OCF} + \frac{\mathrm{T}}{\mathrm{LT}} \, \mathrm{ECF}\label{eq1}
\end{equation}

Embodied carbon can be calculated by quantifying each component of manufacturing emissions~\cite{4}. 
More concretely, the embodied footprint (ECF) is computed by the carbon per unit area (CPA) multiplied by the die area (A), which in turn is dependent on the fab yield (Y), the electrical energy consumed per unit area (EPA), the carbon intensity of the electrical energy (CI), the emissions per unit area from process gases (GPA), and the emissions per unit area associated with raw‑material procurement (MPA), as expressed below:

\vspace{-1.0\baselineskip}
\begin{equation}
\begin{split}
\mathrm{ECF_{SoC}} &= \mathrm{CPA} \times \mathrm{A}  \\
&= \frac{1}{\mathrm{Y}}
\left( \left( \mathrm{CI_{fab}} \times \mathrm{EPA} + \mathrm{GPA} + \mathrm{MPA} \right) \times \mathrm{A} \right)
\end{split}
\label{eq2}
\end{equation}
\vspace{-0.3\baselineskip}

Recent studies show that the carbon footprint of mobile systems has shifted from operational to embodied emissions, directing the focus of sustainable computing research toward the manufacturing stage~\cite{3}. Existing carbon estimation frameworks, including ACT~\cite{4}, 3D-Carbon~\cite{29} and ECO-CHIP~\cite{sudarshan2024eco} for 3D and 2.5D ICs, EPiCarbon~\cite{fayza2025epicarbon}, and nano-3D system~\cite{31}, focus on CMOS or other non-eNVM, existing process flows. 
Their manufacturing energy models rely on CMOS data from iMec~\cite{32}. 
As a result, they give no coverage to the extra and modified stages used in eNVM fabrication without flexibility to incorporate device-specific parameters. 

\subsection{FeFET Technology} 
\label{sec:fefet_background}
\begin{figure}[t]
    \centering    \includegraphics[width=\columnwidth]{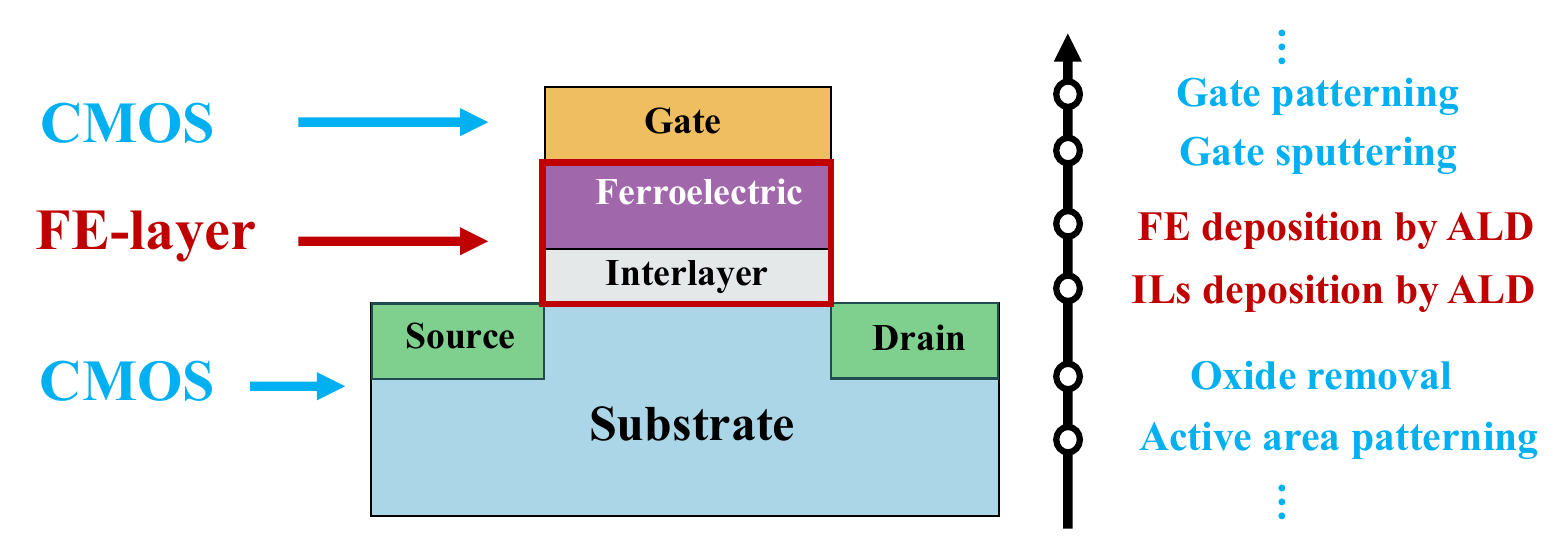} 
    \vspace{-1.5\baselineskip}
    \caption{Take HZO FeFET as an example, the manufacturing steps can be divided into two parts. One is the traditional CMOS baseline, marked by blue color, the other is FEOL special steps to deposit a ferroelectric layer and (optionally) an interfacial layer, bounded by red outline.~\cite{25}.}
    \vspace{-1.5\baselineskip}
    \label{fig2}
\end{figure}

As shown in Figure~\ref{fig2}, FeFETs incorporate a ferroelectric layer into the gate stack, enabling threshold modulation through ferroelectric polarization to encode a stored value~\cite{9}. 
This mechanism provides non-volatility and compact single-transistor bitcells, making FeFETs attractive for high-density on-chip memory with low leakage. 
HZO-based FeFETs have emerged as a highly CMOS-compatible technology, enabling large-scale integration at the advances node~\cite{11}. 
Recent studies have reported HZO-based FeFETs scaled to sub-5\,nm ferroelectric thickness while still exhibiting strong ferroelectricity~\cite{20}. 
This scaling enables polarization switching at relatively low operating voltages, thereby reducing overall write energy and also lowering the demand for peripheral circuits, such as charge pumps, that supply the write voltage.

However, HZO-based FeFETs face critical challenges, notably limited endurance caused by charge trapping~\cite{26}. 
Excessive write operations exacerbate these effects, further degrading device reliability and reducing operational lifetime.

\subsection{Performance modeling tools of eNVMs}
\label{sec:model_limitations}
Several tools are capable of evaluating and optimizing the energy efficiency of eNVM-based systems, with applicability to FeFET-based designs such as HZO-FeFET. 
NVSim~\cite{12} characterizes eNVM array power and area at the circuit level for a particular cell configuration and device type.  
NVMExplorer~\cite{NVMExplorer} offers a cross-stack simulation platform that combines top-conference device databases and application characteristics with underlying characterization via NVSim. 
These tools quantify FeFET power, performance, and area (PPA) and demonstrate its runtime energy advantages as on-chip memory. However, existing tools do not address life cycle sustainability, including the impact of manufacturing.

\section{Proposed Framework}

\begin{figure}[t]
    \centering
    \includegraphics[width=\columnwidth]{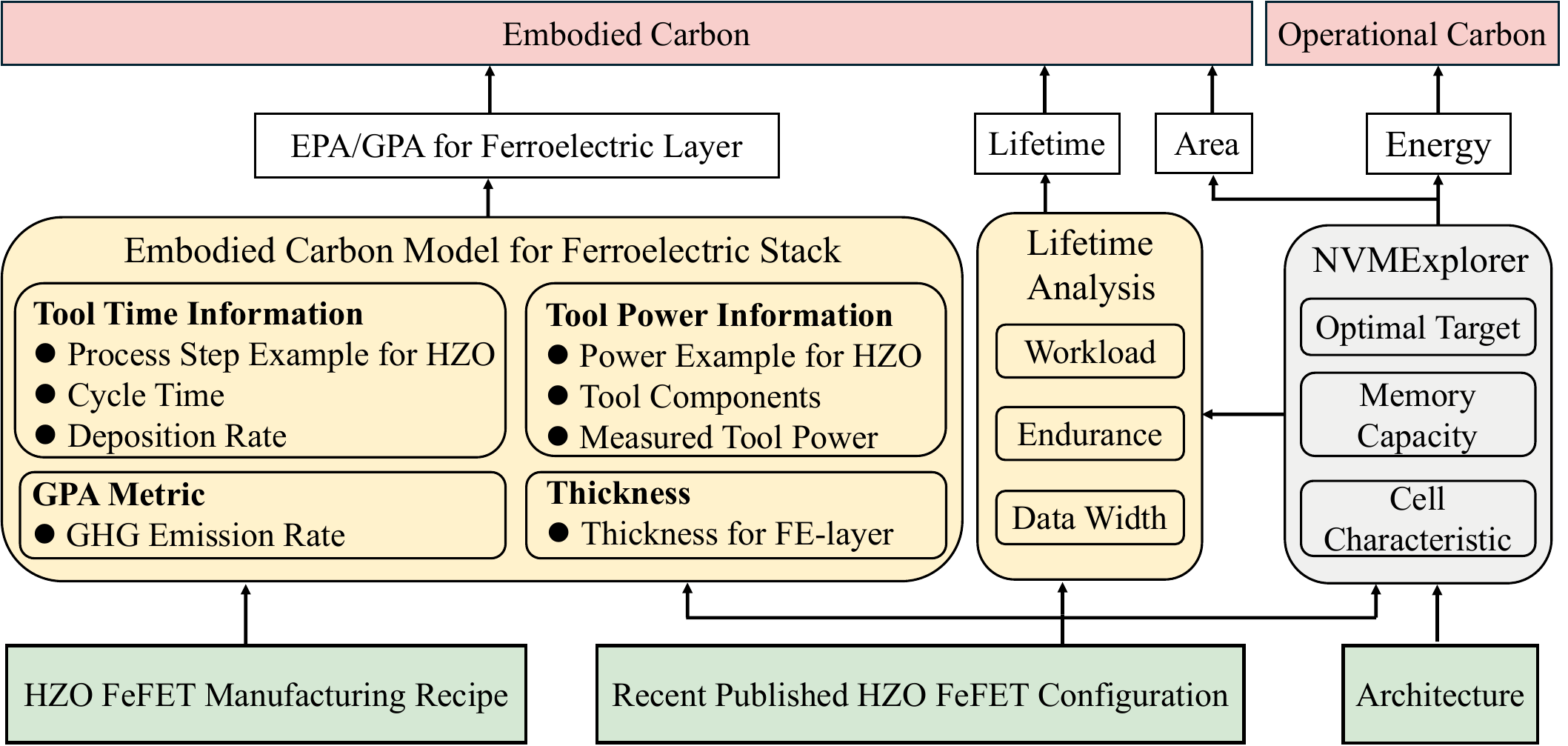} 
    \vspace{-1.3\baselineskip}
    \caption{Diagram of carbon calculation in the proposed COFFEE framework.  Source inputs (green) include  manufacturing recipe and recent published device configurations~\cite{21,22,23,24,25,26,27}. Pink blocks represent carbon categories. 
    Yellow  blocks denote the integrated embodied carbon tools and lifetime analysis, while gray boxes~\cite{NVMExplorer} indicate carbon-related attributes integrated in existing tools.}
    \vspace{-1\baselineskip}
    \label{fig3}
\end{figure}

As shown in Figure~\ref{fig3}, COFFEE is a cross-stack framework that integrates two key new components: (i) an embodied carbon model that estimates the fabrication-stage emissions of HZO-based FeFETs, and (ii) an operational carbon model that derives operational energy eNVM architectural design space exploration tools (i.e., NVMExplorer~\cite{NVMExplorer}). 
Inputs to COFFEE include both manufacturing recipe and HZO FeFET cell characteristics from recent device publications. 
The recipes provide detailed characterization of process steps, including examples of tool power and tool time, while publications supply process thicknesses for embodied carbon evaluation and endurance data for lifetime analysis.
The HZO FeFET cell characteristics are combined with architecture parameters such as memory capacity to form NVMExplorer configurations. NVMExplorer explores design space under user-defined optimal target constraint (e.g, area, read/write latency, memory leakage power, and read/write energy delay product) to identify feasible design points. 
The overall EPA/CPA of HZO-FeFETs (Section~\ref{sec:epa_overall}) is derived from the output parameters EPA/CPA of the ferroelectric layers (Section~\ref{sec:epa_fe_layer}). The EPA/CPA value for FeFET, together with lifetime estimates (Section~\ref{sec:lifetime_analysis}) and runtime energy metrics, are then fed into the carbon models~\cite{4}, to enable end-to-end carbon evaluation of HZO FeFET.

\subsection{Method for FeFET carbon modeling}
\label{sec:epa_overall}

\begin{table}[t]
\scriptsize
\caption{Parameter ranges and sources. The \emph{Source} column indicates origin, as described in Sec~\ref{sec:epa_overall}}

\vspace{-0.8\baselineskip}
\centering
\setlength{\tabcolsep}{3pt}
\renewcommand{\arraystretch}{1.1}
\resizebox{0.8\columnwidth}{!}{
\begin{tabular}{|c|c|c|}
\hline
\textbf{Parameter} & \textbf{Range} & \textbf{Source} \\
\hline
\multicolumn{3}{|c|}{\textbf{ALD manufacturing process flow design related parameters}} \\
\hline
\textnormal{$t_{\mathrm{layer}}$} & 3 (Al$_2$O$_3$), 20 (HZO) nm & Recipe \\
\textnormal{$T_{\mathrm{cycle}}$} & 30 (Al$_2$O$_3$), 100 (HZO) s & Recipe \\
\textnormal{$R_{\mathrm{dep}}$} & 0.1 (Al$_2$O$_3$), 0.2 (HZO) nm/cycle & Recipe \\
\textnormal{$R_{\mathrm{GHG}}$} & 26.9 µg/(nm·cm$^{2}$) & \cite{35} \\
\hline
\multicolumn{3}{|c|}{\textbf{Foundry related parameters}} \\
\hline
Process node & 28 nm & \cite{32} \\
\textnormal{$\mathrm{GPA}_{\mathrm{CMOS}}$} & 0.1375 kg CO$_2$/cm$^{2}$ & \cite{32} \\
\textnormal{$\mathrm{MPA}_{\mathrm{CMOS}}$} & 0.5 kg CO$_2$/cm$^{2}$ & \cite{32} \\
\textnormal{$\mathrm{EPA}_{\mathrm{CMOS}}$} & 0.9 kWh/cm$^{2}$ & \cite{32} \\
\textnormal{$\mathrm{GPA}_{\mathrm{FE\text{-}layer}}$} & 225.96 µg CO$_2$/cm$^{2}$ & \cite{35} \\
\textnormal{$\mathrm{EPA}_{\mathrm{FE\text{-}layer}}$} & 0.26 kWh/cm$^{2}$ & Recipe \\
\hline
\end{tabular}
}
\vspace{-1.6\baselineskip}
\label{tab:parameter_ranges}
\end{table}

The embodied carbon of FeFET-based chips is determined by manufacturing energy consumption, GHG emissions, and facility overheads, which are computed using Equation \eqref{eq2}. 
Table~\ref{tab:parameter_ranges} provides the detailed parameter definitions and the corresponding ranges. 
The \emph{Source} column lists the origin of each parameter: standard CMOS data from iMec~\cite{32}, HZO process flow data from recipe, and GPA metrics from~\cite{35}.

\subsubsection{EPA accounting for FeFET}

To compute the FeFET manufacturing energy, we separate the contributions from the conventional CMOS baseline and the FeFET-specific layers. 
An area-weighted formulation is adopted, where the ferroelectric layer is assumed to be precisely deposited (e.g., the ferroelectric ALD steps) only to the FeFET array region. 
The per-unit-area manufacturing energy of the CMOS baseline ($\mathrm{EPA}_{\mathrm{CMOS}}$) is based on SOTA architectural carbon modeling tools (i.e., ACT~\cite{4}).
Summing the two components yields the overall per-unit-area manufacturing energy of FeFET ($\mathrm{EPA}_{\mathrm{FeFET}}$), where the ratio of the ferroelectric deposition area to the total on-chip memory area is obtained from the reported area efficiency (AE) of cell arrays vs. total array area including CMOS peripherals~\cite{NVMExplorer}:

\vspace{-1.3\baselineskip}
\begin{equation}
\begin{split}
\mathrm{EPA}_{\mathrm{FeFET}} = \mathrm{EPA}_{\mathrm{CMOS}} 
+ \mathrm{EPA}_{\mathrm{FE-layer}} \cdot 
\mathrm{AE}
\end{split}
\label{eq3}
\end{equation}
\vspace{-1.3\baselineskip}

\subsubsection{GPA accounting for FeFET}

Similar to EPA accounting, an area-weighted formulation will be used to integrate the GHG emission for FeFET accounting for the complete on-chip memory:

\vspace*{-1.5\baselineskip}
\begin{equation}
\begin{split}
\mathrm{GPA}_{\mathrm{FeFET}} = \mathrm{GPA}_{\mathrm{CMOS}} 
+ \mathrm{GPA}_{\mathrm{FE-layer}} \cdot 
\mathrm{AE}
\end{split}
\label{eq4}
\end{equation}
\vspace*{-1.4\baselineskip}

\subsubsection{MPA and Yield accounting for FeFET}
For MPA and Yield, the additional ALD steps for HZO-FeFET integration introduce no high-emission precursors and maintain compatibility with standard CMOS processes~\cite{11}. 
Therefore, both MPA and Yield are assumed to be identical to those of the baseline CMOS flow.

\subsection{Embodied carbon model for ferroelectric stack}
\label{sec:epa_fe_layer}
The embodied carbon model integrates tool-time information, tool-power measurements for HZO process steps, GPA metrics, and ferroelectric-layer thickness to compute the GPA and EPA of the ferroelectric layer. 
All parameters are fully configurable, allowing technology designers to substitute their own power values and tool-time data based on alternative fabrication measurements.

\subsubsection{FeFET layer definition}

Compared with a baseline CMOS flow, the HZO-FeFET process adds up to two specialized ferroelectric layers (HZO ferroelectric stack and Al$_2$O$_3$ interface layer) fabricated using ALD technology (Figure~\ref{fig4}).
In practice, the fabrication of the ferroelectric layer also involves an etching step. 
However, since this process is performed together with the gate stack after gate sputtering, its additional energy contribution is negligible compared with that of the baseline CMOS process. 
Therefore, the analysis only requires computing the manufacturing energy of the two ferroelectric layers during the ALD process.

\subsubsection{Process steps for HZO layers}

\begin{figure}[t]
    \centering
    \includegraphics[width=\columnwidth]{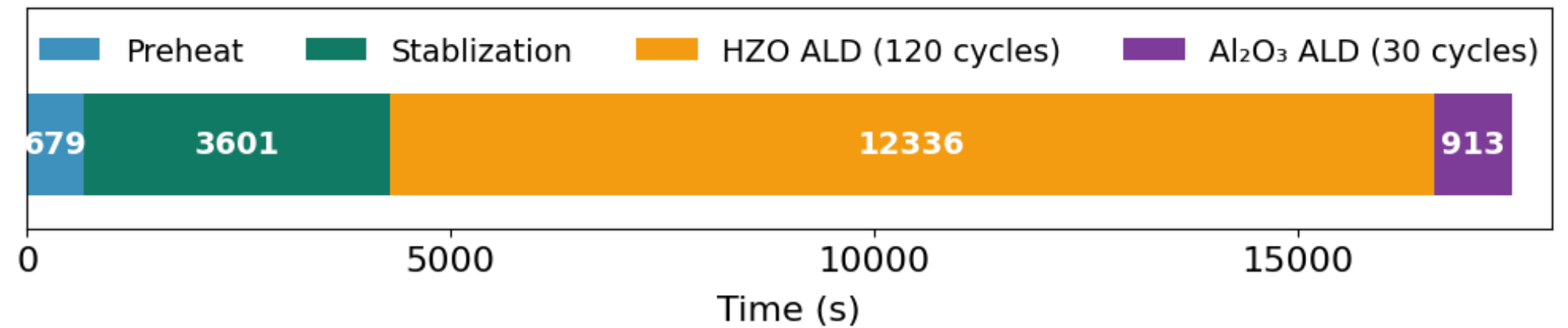} 
    \vspace*{-1.7\baselineskip}
    \caption{The ALD process is divided into four stages: pre-heat, stabilization, HZO deposition, and Al$_2$O$_3$ deposition. For a 20\,nm HZO layer and a 3\,nm Al$_2$O$_3$ layer, the total process time reaches 17529\,s.
}
    \vspace*{-1.2\baselineskip}
    \label{fig4}
\end{figure}

\begin{figure}[t]
    \centering
    \includegraphics[width=\columnwidth]{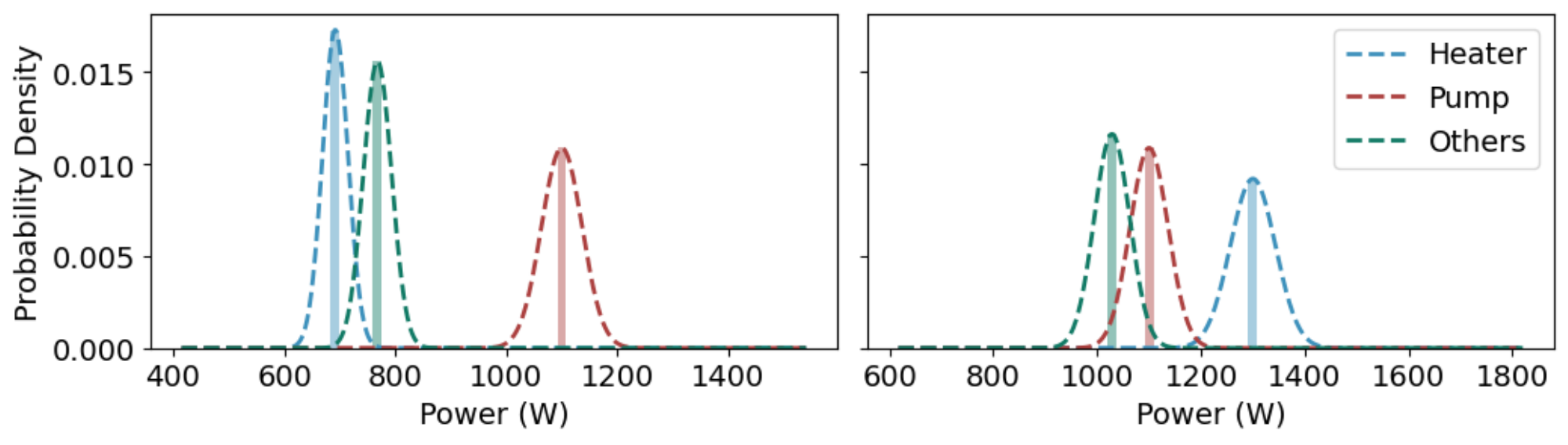} 
    \vspace{-2.0\baselineskip}
    \caption{Power consumption of fabrication equipment, including the chamber, pump, and other modules. The left plot shows the equipment power during the heating stage ($P_{\mathrm{preheat}}$) with uncertainty distribution~\cite{carboniccad}. The right plot shows the equipment power during steady operation at 200$\,^{\circ}\mathrm{C}$ ($P_{\mathrm{steady}}$).}
    \vspace{-1.5\baselineskip}
    \label{fig5}
\end{figure}

The ALD process for HZO layers followed recipe specifying a 20\,nm HZO layer and a 3\,nm Al$_2$O$_3$ layer. Deposition is partitioned into four stages: preheat, stabilization, HZO deposition, and Al$_2$O$_3$ deposition (Figure~\ref{fig4}). 
A deposition temperature of 200 °C was chosen for endurance and BEOL compatibility.

\subsubsection{EPA for FE-layer}

The manufacturing energy consumption for each stage is calculated by multiplying the ALD tool's power ($P_n$) shown in Fig.~\ref{fig5} by its operating time ($T_n$). 
Summing across all four stages yields the FeFET-layer manufacturing energy by tool. 
In addition to tool power, we account for facility support by applying a 40\% overhead ($f_{\mathrm{facility}}$) following~\cite{32}.
Dividing the total manufacturing energy by the wafer area($A_{\mathrm{Wafer}}$) produces the EPA for the FeFET layer:

\vspace{-0.6\baselineskip}
\begin{equation}
\mathrm{EPA}_{\mathrm{FE\mathchar`-layer}} = 
\frac{\sum \left( T_n \cdot  P_n \right)}
{A_{\mathrm{Wafer}} \cdot  (1 - f_{\mathrm{facility}})}
\label{eq5}
\end{equation}
\vspace{-1.0\baselineskip}

The deposition processes of HZO and Al$_2$O$_3$ are configurable and determined by specific ALD parameters together with the target ferroelectric layer thickness ($t_{\mathrm{layer}}$). 
The ALD parameters include the deposition rate ($R_{\mathrm{dep}}$), which denotes the film thickness deposited per cycle, and the cycle time ($T_{\mathrm{cycle}}$), which denotes the duration of a single cycle~\cite{35}. 
The deposition time for each layer can thus be expressed as:
\vspace{-0.5\baselineskip}
\begin{equation}
T_{\mathrm{deposition}} = \frac{t_{\mathrm{layer}}}{R_{\mathrm{dep}}} \cdot T_{\mathrm{cycle}}
\label{eq6}
\end{equation}
\vspace{-0.8\baselineskip}

To model power consumption during fabrication, we decompose the ALD tool into two dominant components: (i) the chamber, which provides the reaction environment and whose power is governed by the heater setpoint, and (ii) the pump, which maintains chamber pressure and transports gases at approximately constant power. During preheat, the chamber is pumped down, heaters are set to deposition temperature.
As shown in Figure~\ref{fig5}, the chamber ramps to 200$\,^{\circ}\mathrm{C}$ and draws $P_{\mathrm{preheat}}$, while in subsequent stages it operates near 200$\,^{\circ}\mathrm{C}$ with steady power $P_{\mathrm{steady}}$. We attribute 70\% of the tool power to the chamber and pump, with the remaining 30\% assigned to other modules such as control electronics~\cite{39}. To account for uncertainty and fluctuations around average (nominal) power, each nominal value is perturbed by a Gaussian distribution, with three-sigma set to 10\% of the nominal power~\cite{carboniccad}.

\begin{table}[t] \caption{Device Parameters for HZO1 to HZO7} \centering \scriptsize \setlength{\tabcolsep}{4pt} \renewcommand{\arraystretch}{1.1} 
\vspace{-0.8\baselineskip}
\resizebox{\columnwidth}{!}{ \begin{tabular}{|c|c|c|c|c|c|c|c|} \hline \textbf{Metric} & \textbf{HZO1} & \textbf{HZO2} & \textbf{HZO3} & \textbf{HZO4} & \textbf{HZO5} & \textbf{HZO6} & \textbf{HZO7} \\ 
\hline 
Read\_voltage (V) & 0.1 & 0.9 & 0.1 & 0.1 & 0.1 & 0.1 & 0.1 \\ 
Reset\_voltage (V) & -3 & -4.5 & -5 & 8 & 8 & -4 & -3.5 \\ 
Set\_voltage (V) & 3 & 7.5 & 5 & -6 & -6 & 4 & 4.5\\
Program pulse (ns)  & 10000 & 10000 & 100 & 100 & 50 & 2000 & 1000 \\
Endurance (cycles) & $10^{5}$ & $10^{4}$ & $10^{5}$ & $10^{9}$ & $10^{9}$ & $10^{6}$ & $10^{5}$ \\
HZO thickness (nm) & 10 & 10 & 15 & 16.35 & 16.35 & 12.71 & 10 \\ 
Al$_2$O$_3$ thickness (nm) & 3 & 3 & 0 & 2.64 & 2.68 & 0 & 1 \\ 
Reference & ~\cite{21} & ~\cite{22} & ~\cite{23} & ~\cite{24} & ~\cite{25} & ~\cite{26} & ~\cite{27}\\ \hline 
\end{tabular}} 
\vspace{-1\baselineskip}
\label{tab:hzo_params} \end{table}

\begin{figure}[t]
    \centering
    \includegraphics[width=\columnwidth]{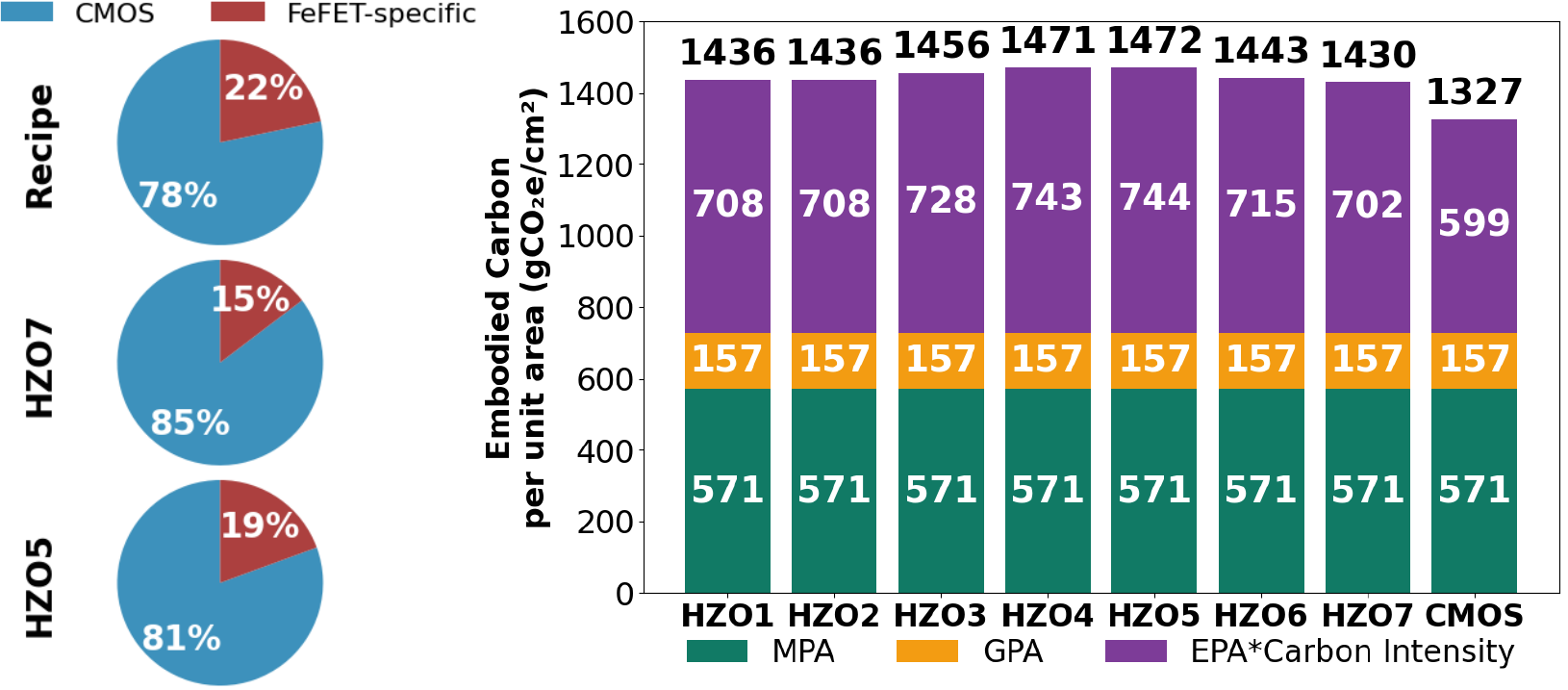} 
    \vspace{-1.5\baselineskip}
    \caption{(Left) EPA breakdown for 2MB HZO-FeFETs across HZO5, HZO7 and recipe data (20\,nm HZO and 3\,nm Al$_2$O$_3$). (Right) CPA breakdown for 2MB HZO FeFET(HZO1-HZO7) and CMOS baseline. Adding ferroelectric layers contributes about 20\% of the total EPA in HZO-FeFETs, leading to an $\sim$10\% increase in overall CPA.}
    \vspace{-1.5\baselineskip}
    \label{fig6}
\end{figure}

\subsubsection{GPA for FE-layer}

Following the iMec methodology for GHG accounting, the gas mass $m_i$ denotes the amount of GHGs emitted by processing tools~\cite{32}. The global warming potential (GWP) per unit mass and the destruction efficiency $d_i$ of abatement systems are obtained from~\cite{37,38}.

\vspace{-0.8\baselineskip}
\begin{equation}
\mathrm{GPA}_{\mathrm{FE-layer}}
= \mathrm{GWP}\cdot \sum m_{i}\cdot \bigl(1 - d_{i}\bigr)
\label{eq7}
\end{equation}
\vspace{-1.2\baselineskip}

\begin{figure*}[t]
    \centering
    \includegraphics[width=\textwidth]{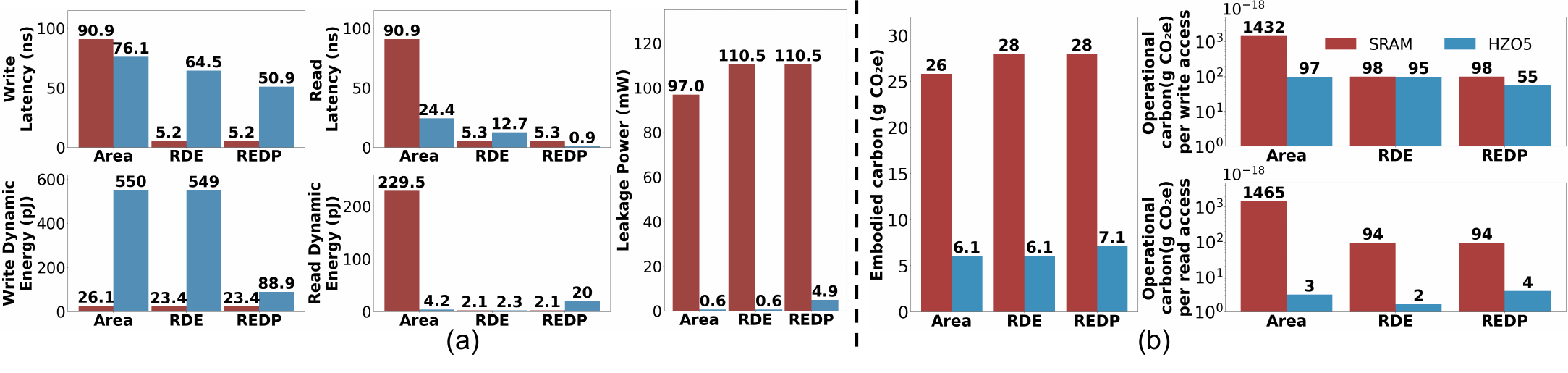} 
    \vspace{-2.3\baselineskip}
    \caption{ (a) Key performance metrics for 2\,MB HZO5 FeFET and SRAM under different optimal targets (RDE: ReadDynamicEnergy, REDP: ReadEDP). 
    (b) Carbon characteristics of the same configurations: embodied carbon and per-access operational carbon for write and read operation.Compared to SRAM, HZO-FeFETs achieve lower embodied and operational carbon, primarily due to smaller cell area and reduced leakage power.}
    \vspace{-1.6\baselineskip}
    \label{fig7}
\end{figure*}

For GHG accounting, only methane (CH$_4$) is emitted in small quantities during the Al$_2$O$_3$ ALD deposition step. 
As shown in~\cite{29}, the mass of CH$_4$ produced, $m_{\mathrm{CH_4}}$, is proportional to the mass of the deposited film. 
By extracting the CH$_4$ mass and deposition rate from~\cite{29}, the GHG emission rate ($R_{\mathrm{GHG}}$), defined as the CH$_4$ mass per unit film thickness, can be obtained. 
Multiplying this rate by the Al$_2$O$_3$ film thickness ($t_{\mathrm{Al_2O_3}}$) yields the CH$_4$ mass per unit area ($m_{\mathrm{CH_{4}}}$):
\vspace{-0.7\baselineskip}
\begin{equation}
m_{\mathrm{CH_4}} = R_{\mathrm{GHG}} \cdot t_{\mathrm{Al_2O_3}}
\label{eq:co2eq}
\end{equation}

\subsection{Lifetime analysis}
\label{sec:lifetime_analysis}
To estimate lifetime, we assume continuous operation at a specified write access rate or application use case, as in prior work~\cite{NVMExplorer}. 
The limiting factor is the intrinsic endurance of varying HZO FeFET devices, as reported in recent publications~\cite{21,22,23,24,25,26,27}. 
For a lifetime estimate assuming write traffic distributed across the entire capacity $C_{\mathrm{mem}}$ of the memory array, we use potential write traffic ($t_{\mathrm{write}}$), i.e., the number of write operations per day, and the size of each access, i.e., the array data width ($W_{\mathrm{data}}$), to estimate the number of days of operation before reliability is expected to degrade:
\vspace{-0.2\baselineskip}
\begin{equation}
\mathrm{Lifetime} 
= \frac{\mathrm{Endurance} \cdot C_{\mathrm{mem}}}
{t_{\mathrm{write}} \cdot W_{\mathrm{data}}}
\label{eq:lifetime}
\end{equation}

\section{Evaluation}

In this section, we apply the COFFEE framework to evaluate the EPA and CPA for reported HZO-FeFETs (Sec.~\ref{sec:carbon_breakdown}), compare an example HZO-FeFET with SRAM to analyze the trade-off between carbon and energy efficiency (Sec.~\ref{sec:carbon_vs_energy}), and finally discuss the impact of device lifetime (Sec.~\ref{sec:lifetime}).

\subsection{Result for EPA and CPA}
\label{sec:carbon_breakdown}

\textbf{Takeaway 1:} \textit{Adding the ferroelectric layers contributes roughly 20\% of total EPA in HZO-FeFETs, resulting in up to an 11\% increase in overall CPA.
}

We consider seven HZO-based FeFETs (HZO1–HZO7) reported in recent publications~\cite{21,22,23,24,25,26,27}. Publications do not report an exact cell size, so a consistent value of 30$F^{2}$ is assumed for the devices according to previously reported FeFET characteristics~\cite{8}. 
The corresponding device parameters and process layer thicknesses are summarized in Table~\ref{tab:hzo_params}.

Figure~\ref{fig6} (left) shows the EPA breakdown of 2\,MB HZO7 FeFETs (thinnest thickness) and HZO5 FeFETs (thickest thickness), together with fabrication recipe data (Table~\ref{tab:parameter_ranges}), with an array architecture optimized for minimum area. 
In these cases, the ferroelectric layer contributes approximately 20\% of the total EPA.  
Figure~\ref{fig6} (right) compares CPA between 2\,MB HZO-FeFETs and CMOS under the \emph{Area} target (using Taiwan’s carbon intensity of 583\,g/kWh~\cite{4}). The breakdown values for both CMOS and HZO FeFET are normalized by yield (Y = 0.875). 
Because the GPA from the ferroelectric layer at nominal thickness (3\,nm) yields only microgram-level CO$_2$ emissions, the overall GPA of FeFETs is nearly identical to CMOS. 
By contrast, extreme ferroelectric thickness (HZO5 FeFET) can contribute up to 11\% additional CO$_2$ emissions.

\subsection{Carbon result for HZO FeFET}
\label{sec:carbon_vs_energy}

\textbf{Takeaway 2:} \textit{Compared to SRAM, the HZO FeFET achieves lower embodied and operational carbon, primarily due to its smaller cell area and reduced leakage power.}

To analyze the design space between sustainability and energy efficiency, we compare SRAM (cell size = 146$F^{2}$) against the HZO5 FeFET, which exhibits the highest CPA and the lowest write latency among seven devices. Figure~\ref{fig7}(a) illustrates the performance metrics for 2\,MB SRAM and HZO5 FeFET, while Figure~\ref{fig7}(b) presents  embodied carbon and operational carbon per access for the two memories. 
Energy, latency, and energy-per-access metrics are obtained using NVMExplorer, and operational carbon per access is then estimated by combining the energy-per-access results with Taiwan’s carbon intensity (583\,g/kWh~\cite{4}).
We optimize the array architecture (i.e., provisioning dimensions and peripherals) under three targets: minimum \emph{Area}, minimum \emph{Read Dynamic Energy} (RDE), and minimum  \emph{Read Energy-Delay Product} (REDP).

Figure~\ref{fig7}(a) shows that SRAM is write-efficient with lower latency and dynamic energy, but suffers from leakage power up to 179$\times$ higher than HZO5 FeFET. 
Among the optimal targets, \emph{Area} yields the lowest leakage power by minimizing peripheral circuitry, while \emph{REDP} minimizes access latency.

Figure~\ref{fig7}(b) shows that the smaller cell size of HZO5 reduces the array footprint, and despite its additional ferroelectric stack increasing CPA, its overall embodied footprint remains lower than that of SRAM. In addition, HZO5 achieves consistently lower per-access operational carbon across optimization targets, driven primarily by its significantly lower leakage power, with average reductions of 186$\times$ for read and 5.85$\times$ for write.

Overall, SRAM is write latency-optimal, while HZO5 is carbon-optimal. 
In system settings when write access latency must remain below 10\,ns, only SRAM meets the requirement, but at a higher carbon cost. 
Otherwise, HZO5 is preferable to minimize both embodied and operational carbon: \emph{Area} is most effective when embodied carbon is prioritized, \emph{REDP} balances carbon and latency, and \emph{RDE} minimizes read-side operational carbon with comparable write-side cost.

\begin{figure}[t]
    \centering    \includegraphics[width=\columnwidth]
    {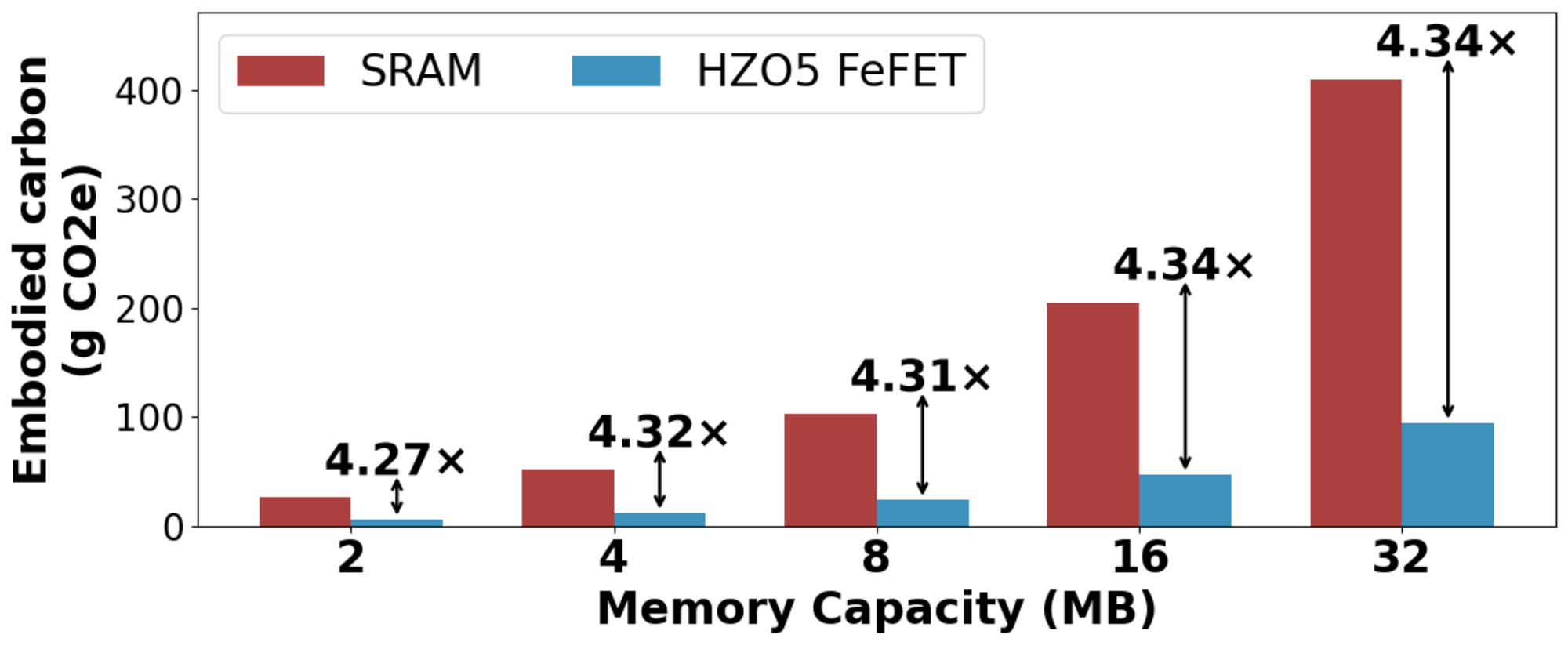} 
    \vspace{-1.8\baselineskip}
    \caption{Embodied carbon of HZO5 FeFET and SRAM across memory capacities under the \emph{Area} target. The SRAM/HZO5 gap remains essentially constant at $\sim4.3\times$.}
    \vspace{-1.5\baselineskip}
    \label{fig8}
\end{figure}

\textbf{Takeaway 3:} \textit{Embodied carbon scales roughly linearly with capacity, with a consistent reduction of about $\sim$4.3$\times$ between SRAM and HZO-FeFET due to the increased storage density despite the increase in carbon-per-unit area.}

We use the \emph{Area}-optimal point as the baseline for capacity scaling because, at 2\,MB, it yields the lowest embodied carbon for both SRAM and HZO5. 
By design, \emph{Area} consistently minimizes peripheral overhead across capacities, so scaling trends primarily reflect array growth rather than array architecture variation. 
Figure~\ref{fig8} shows that embodied carbon increases approximately linearly with capacity for both on-chip memories, while the SRAM–HZO5 gap remains essentially constant due to the smaller cell size and overall improvement in storage density of HZO-5 despite higher CPA.

\subsection{Lifetime discussion}
\label{sec:lifetime}
We focus on write-induced wear as the dominant reliability concern~\cite{43}. 
For HZO-FeFETs, reported devices span approximately $10^{4}$–$10^{9}$ program/erase cycles per cell (Table~\ref{tab:hzo_params}). 
According to Eq.~(\ref{eq:lifetime}), with device type and memory capacity fixed, lifetime scales inversely with write intensity. 
Within this context, HZO5 illustrates a clear trade-off: the thicker ferroelectric stack increases the embodied footprint per area (higher CPA, Figure \ref{fig6}), yet its higher endurance extends the lifetime under the same write load. 
As write intensity rises, the lifetime advantage amortizes the larger upfront footprint, reducing the embodied impact per access. 
In summary, device selection should co-optimize endurance and CPA against the expected write intensity and usable capacity, rather than maximizing either metric in isolation.

\section{Case Study}

\begin{figure}[t!]
    \centering   
    \includegraphics[width=\columnwidth]{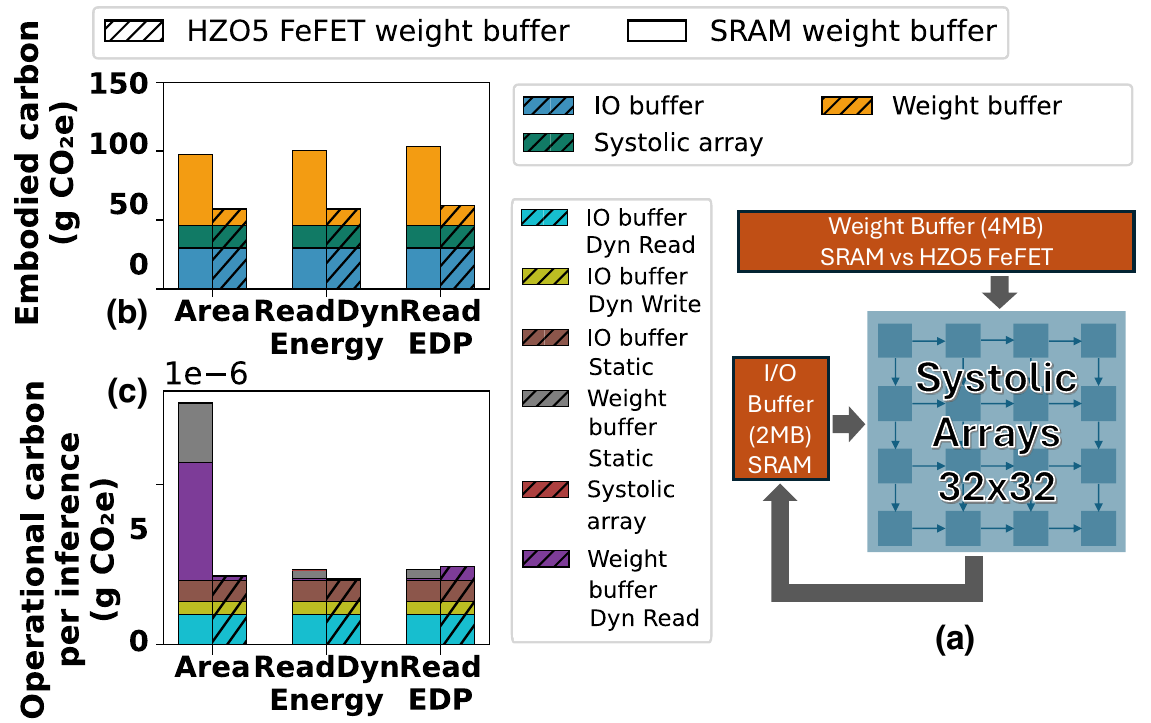}
    \vspace{-8mm}
    \caption{(a) Edge TPU architecture with a 32×32 systolic array and 2 MB IO buffer and 4 MB weight buffer. Embodied (b) and operational (c) carbon per inference for HZO5 FeFET and SRAM under different weight buffer OPT target running MobileNet V1~\cite{mobilenetv1}. We keep the design for SRAM-based IO buffer the same in all scenarios. HZO5-based weight yields significant improvement in terms of embodied carbon and demonstrate competitive advantage in reducing operational carbon. }
    \vspace{-1.2\baselineskip}
    \label{edgeTPUcarbon}
\end{figure}

In this section, we investigate the carbon emission benefits of integrating HZO5 to the weight buffer in the systolic array of an edge TPU~\cite{edgeTPUArchitecture}, leveraging the compact area and low read-latency advantages of FeFET-based eNVM. 
We evaluate MobileNet V1~\cite{mobilenetv1} on a 32×32 systolic array with a 2 MB unified input and output buffer and a 4 MB weight buffer, as shown in Fig.~\ref{edgeTPUcarbon}(a). 

The edge TPU configuration follows the default accelerator model in SCALE-Sim\cite{SCALESIM} and edge TPU architecture described in~\cite{edgeTPUArchitecture}. 
We keep all other configuration of the edge TPU the same and compare the embodied and operational (per inference) of a SRAM-based vs HZO5-based weight buffer. 
The goal is to exploit HZO5’s low read latency for read-heavy traffic; additionally, the limited write rate to the weight buffer helps extend HZO5 lifetime.

To estimate the carbon footprint of the edge TPU under different weight buffer implementations, we use NVMExplorer~\cite{NVMExplorer} to determine buffer areas, which are then fed into COFFEE to compute embodied carbon emission. 
The operational carbon is computed by multiplying the number of access to each buffer with dynamic energy per access and the execution time multiplied by leakage power. 
The number of buffer accesses and execution time are based on SCALE-Sim\cite{SCALESIM} simulation of MobileNet V1~\cite{mobilenetv1}, whereas the dynamic energy per access and leakage power are calculated based on NVMExplorer~\cite{NVMExplorer} under different optimization targets. 
We calculate the area and the energy of systolic array based on synthesis data reported in 3D IC~\cite{3D-IC}.
The SRAM IO buffer's carbon impacts are calculated under optimization target of write energy-delay-product, which we observe to demonstrate the overall lowest embodied and operational carbon.

Fig. \ref{edgeTPUcarbon} shows that edge TPU with HZO5-based weight buffer effectively reduce carbon emission: 
HZO5-based weight buffer reduced the total edge TPU architecture area by 49.8\%, which translates to 42.3\% improvement in embodied carbon. 
HZO5 also demonstrates competitive advantage on reducing the operational carbon emission, with 70\% operational carbon per inference improvement compare to SRAM-based weight buffer under area optimized target, and yield the lowest operational carbon per inference under read-dynamic-energy-optimized target across all technologies and target optimization.


\section{Conclusion}

We propose COFFEE, a modeling framework for evaluating HZO-based FeFET eNVMs from a sustainability perspective. 
Our results show that, at 2\,MB capacity, the embodied carbon per unit area overhead of HZO-FeFETs can be up to 11\% higher than the CMOS baseline, while the embodied carbon per MB remains consistently about 4.3$\times$ lower than SRAM under different memory capacity.
We further apply the framework to integrate HZO FeFET to an AI accelerator, achieving up to 42.3\% embodied and 70\% operational carbon reduction. 

\section*{Acknowledgment}

This research was supported by the National Science Foundation under award numbers CCF-2324860 and CNS-2335795.

\bibliographystyle{IEEEtran} 
\clearpage
\bibliography{reference}

\end{document}